\documentclass[%
 reprint,
longbibliography,
superscriptaddress,
 amsmath,amssymb,
 aps,
prb,
]{revtex4-1}

\usepackage{graphicx}% Include figure files
\usepackage{dcolumn}% Align table columns on decimal point
\usepackage{bm}% bold math
\usepackage{float}
\usepackage{bbold}
\usepackage{latexsym}    
\usepackage{wasysym}
\usepackage{amsfonts}
\usepackage{siunitx}
\usepackage{blindtext}
\usepackage{caption} 
\newcommand{\dg}[1]{\ensuremath{#1^\circ}}
\newcommand{\wavenumber}[1]{\ensuremath{#1~\text{cm}^{-1}}}
\newcommand{\mumetr}[1]{\ensuremath{\SI{#1}{\micro\meter}}}
\newcommand{\nmetr}[1]{\ensuremath{#1~\text{nm}}}

\newcommand{\kvec}{\ensuremath{\vec{k}}}
\newcommand{\etal}[1]{#1 \textit{et al}.}
\newcommand{\AMa}[1]{\ensuremath{\bold{A}_{#1}}}
\newcommand{\PMa}[1]{\ensuremath{\bold{P}_{#1}}}
\newcommand{\TMa}[1]{\ensuremath{\bold{T}_{#1}}}
\newcommand{\LMa}[1]{\ensuremath{\bold{L}_{#1}}}

\newcommand{\epsiTens}{\ensuremath{\bar{\varepsilon}}}
\newcommand{\incAngle}{\ensuremath{\theta}}

\newcommand{\epsi}{\ensuremath{\varepsilon}}

\newcommand{\qij}{\ensuremath{q_{ij}}}

\newcommand{\colvec}[1]{\ensuremath{\begin{pmatrix}#1\end{pmatrix}}}

\newcommand{\Del}{\ensuremath{\bold{\Delta}}}

\newcommand{\abs}[1]{\ensuremath{\lvert #1 \rvert}} 
\newcommand\T{\ensuremath{\mathcal{T}}}
\newcommand\R{\ensuremath{\mathcal{R}}}
\newcommand\A{\ensuremath{\mathcal{A}}}
\newcommand\E{\ensuremath{\mathcal{E}}}
\newcommand\HH{\ensuremath{\mathcal{H}}}

\newcommand\SSS{\ensuremath{\mathcal{S}}}
\newcommand{\Hvec}{\ensuremath{\vec{H}}}

\newcommand{\Evec}{\ensuremath{\vec{E}}}
\newcommand{\Svec}{\ensuremath{\vec{S}}}

\newcommand{\FullTMa}[1]{\ensuremath{\bold{\Gamma}_{#1}}}
\newcommand{\FullTMaEl}[1]{\ensuremath{\Gamma}_{#1}}

\newcommand{\MoO}{\ensuremath{\text{MoO}_{\text{3}}}}
\newcommand{\MoS}{\ensuremath{\text{MoS}_{\text{2}}}}
\newcommand{\WSe}{\ensuremath{\text{WSe}_{\text{2}}}}
\newcommand{\WS}{\ensuremath{\text{WS}_{\text{2}}}}
\newcommand{\SiO}{\ensuremath{\text{SiO}_{\text{2}}}}
\newcommand{\w}[2]{\ensuremath{\omega_{\text{#1}}^{\text{#2}}}}

\newcommand{\fourfourMatrix}[4]{\ensuremath{\begin{pmatrix}
#1 \\
#2 \\ 
#3 \\
#4 \end{pmatrix}}}

\begin{document}

%\preprint{APS/123-QED}

\title{Layer-Resolved Absorption of Light in Arbitrarily Anisotropic Heterostructures}% Force line breaks with \\

\author{Nikolai Christian Passler}
\email{passler@fhi-berlin.mpg.de}
\affiliation{Fritz-Haber-Institut der Max-Planck-Gesellschaft, Faradayweg 4-6,14195 Berlin, Germany}
\author{Mathieu Jeannin}
\affiliation{Laboratoire de Physique de l'\'{E}cole Normale Sup\'{e}rieure, ENS, Paris Sciences Lettres, CNRS, Universit\'{e} de Paris, 24 Rue Lhomond, 75005 Paris, France}
\author{Alexander Paarmann}
\email{alexander.paarmann@fhi-berlin.mpg.de}
\affiliation{Fritz-Haber-Institut der Max-Planck-Gesellschaft, Faradayweg 4-6,14195 Berlin, Germany}

\date{\today}% It is always \today, today,

\begin{abstract}
We present a generalized formalism to describe the optical energy flow and spatially resolved absorption in arbitrarily anisotropic layered structures. The algorithm is capable of treating any number of layers of arbitrarily anisotropic, birefringent, and absorbing media and is implemented in an open access computer program. We derive explicit expressions for the transmitted and absorbed power at any point in the multilayer structure, using the electric field distribution from a $4 \times 4$ transfer matrix formalism. As a test ground, we study three nanophotonic device structures featuring unique layer-resolved absorption characteristics, making use of (i) in-plane hyperbolic phonon polaritons, (ii) layer-selective, cavity-enhanced exciton absorption in transition metal dichalcogenide monolayers, and (iii) intersubband-cavity polaritons in quantum wells. Covering such a broad spectral range from the far-infrared to the visible, the case studies demonstrate the generality and wide applicability of our approach. 
%\begin{description}
%%\item[Usage]
%%Secondary publications and information retrieval purposes.
%%\item[PACS numbers]
%%May be entered using the \verb+\pacs{#1}+ command.
%%\item[Structure]
%%You may use the \texttt{description} environment to structure your abstract;
%%use the optional argument of the \verb+\item+ command to give the category of each item. 
%\end{description}
\end{abstract}

\maketitle

%\tableofcontents

\section{Introduction}
The absorption of light in thin layers of strongly anisotropic materials has received enormous attention since the dawn of two-dimensional (2D) materials and their heterostructures \cite{Geim2013}. Tremendous progress has been reported using 2D materials for numerous nanophotonic applications such as hybrid graphene-based photodetectors \cite{Gabor2011,Freitag2013}, optoelectronic \cite{He2013,Ross2014} and photovoltaic \cite{Fortin1982,Yu2013} devices employing transition metal dichalcogenide (TMDC) monolayers, enhanced light-matter interaction using photonic integration with optical cavities \cite{Liu2011,Furchi2012}, approaches towards TMDC-based nanolasers \cite{Gan2013,Sobhani2014}, hyperlensing \cite{Dai2015,Ferrari2015} based on hyperbolic polaritons \cite{Ma2018}, bio-sensing \cite{Rodrigo2015}, and thermoelectric applications using black phosphorus (BP) monolayers \cite{Fei2014,Xia2014}. In light of these thriving developments and the great potential entailed in nanophotonic technology, a robust and consistent theoretical framework for the description of light-matter interaction in layered heterostructures of anisotropic materials is of central importance.
%Modern nanophotonics is driven by the enhanced light-matter interaction in complex hybrid nanostructures \cite{Koenderink2015}. In particular polaritons, photons that strongly couple to a material's quasiparticle, have proven to enable immense electric field strengths localized on lengthscales far below the refraction limit \cite{Hillenbrand2002,Gramotnev2010}. Polaritons in ultra-thin films, monolayers, or stratified heterostructures constitute well-defined and easily manufacturable systems, and thus have been extensively studied, giving rise to numerous applications such as sensing \cite{Neuner2010,Sinibaldi2012}, opto-electronic control \cite{Eda2013,Zhao2015}, ultrafast switching \cite{Li2016,Dunkelberger2018}, or hyperlensing \cite{Dai2015,Ferrari2015}. 
%5690

In order to understand, analyze and predict the optical response of multilayer structures, the transfer matrix formalism has proven to be of great utility \cite{Scarlat2010,Mounier2011,Ratchford2019}. In isotropic layered media, a $2 \times 2$ transfer matrix fully describes any light-matter interaction, and with knowledge of the local electric and magnetic fields, the optical power flow can be described by the Pointing vector~\SSS~\cite{Chilwell1984,Deparis2011}. However, what already proves intricate in isotropic multilayers, becomes even more cumbersome when the materials are uniaxial or even biaxial requiring a $4 \times 4$ transfer matrix formalism \cite{Berreman1972,Yeh1979,Passler2017a}, as it is the case for many state-of-the-art nanophotonic materials like hexagonal boron nitride (hBN) \cite{Geick1966}, molybdenum trioxide (\MoO) \cite{Lajaunie2013}, or BP \cite{Liu2017}. In consequence, to the best of our knowledge, previous approaches aiming at the analytical computation of light absorption in anisotropic multilayers are restricted to special cases \cite{Collett1971,GiaRusso1973,Schwelb1986,Ciumac1994}, whereas a fully generalized formalism applicable to any number of layers of media with arbitrary permittivity has not been proposed so far.

In this work, we derive explicit expressions for the layer-resolved transmittance and absorption in stratified heterostructures of arbitrarily anisotropic, birefringent, and absorbing media, using the electric field distribution provided by our previous transfer matrix formalism \cite{Passler2017a,Passler2019b}. Our algorithm is numerically stable, yields continuous solutions, and is implemented in an open access computer program \cite{Passler2020,Jeannin2020}, enabling a robust and consistent framework that is capable of treating light of any polarization impinging at any incident angle onto any number of arbitrarily anisotropic, birefringent, and absorbing layers. To demonstrate the capabilities of our algorithm, we present and discuss simulation results for three nanophotonic device structures, featuring several phenomena such as azimuth-dependent hyperbolic phonon polaritons in a \MoO~/ aluminum nitride (AlN) / silicon carbide (SiC) heterostructure, layer-selective exciton absorption of molybdenum disulfide (\MoS) monolayers in a Fabry-P\'{e}rot cavity, and strong light-matter coupling between a cavity mode and an epsilon-near-zero mode in a doped gallium nitride (GaN) multi-quantum well system. Section \ref{sec:theory} summarizes the transfer matrix framework that is used to calculate the electric field distribution and the momenta of the eigenmodes in an anisotropic multilayer system. Based on this theory, section \ref{sec:layer_res_trans_abs} introduces the calculation of the layer-resolved transmittance and absorption. In section \ref{sec:simulations}, the simulation results are presented.

\section{Tranfer Matrix Framework}
\label{sec:theory}
The $4\times 4$ transfer matrix formalism comprising the calculation and sorting of the eigenmodes and the treatment of singularities (Section \ref{sec:matrix_formalism}), the calculation of reflection and transmission coefficients (Section \ref{sec:rt_coefficients}) and of the electric fields (Section \ref{sec:Efield_distribution}) is based on our previous work \cite{Passler2017a}, and therefore is here only briefly summarized in order to provide the necessary framework for the calculation of the layer-resolved absorption (Section \ref{sec:layer_res_trans_abs}).

\subsection{Matrix Formalism}
\label{sec:matrix_formalism}

The incident medium is taken to be non-absorptive with isotropic (relative) permittivity $\epsi_{\text{inc}}$, while all other media can feature absorption and fully anisotropic (relative) permittivity tensors \epsiTens. Each permittivity tensor $\epsiTens_i$ of medium $i$ with principle relative permittivities in the crystal frame $\epsi_x$, $\epsi_y$, and $\epsi_z$ can be rotated into the lab frame using a three-dimensional coordinate rotation matrix \cite{Yeh1979} (Eq.~2 in Ref.~\cite{Passler2017a}). In the following, media with a diagonal permittivity tensor in the lab frame are referred to as non-birefringent, while media with a permittivity tensor that features non-zero off-diagonal elements is called birefringent. Furthermore, all media are assumed to have an isotropic magnetic permeability~$\mu$. 

The coordinate system in the lab frame is defined such that the multilayer interfaces are parallel to the $x$-$y$-plane, while the $z$-direction points from the incident medium towards the substrate and has its origin at the first interface between incident medium and layer $i=1$. The layers are indexed from $i=1$ to $i=N$, and the thickness of each layer is $d_i$. Furthermore, $i=0$ refers to the incident medium and $i=N+1$ to the substrate. The plane of incidence is the $x$-$z$-plane, yielding the following wavevector $\kvec_i$ in layer $i$:
\begin{align}
\kvec_i=\frac{\omega}{c}(\xi,0,q_i),
\label{eq:kvec}
\end{align}
where $\omega$ is the circular frequency of the incident light, $c$ is the speed of light in vacuum, $\xi=\sqrt{\epsi_{\text{inc}}} \: \sin(\theta)$ is the in-plane $x$-component of the wavevector which is conserved throughout the entire multilayer system, \incAngle~is the incident angle, and $q_i$ is the dimensionless $z$-component of the wavevector in layer $i$. 

%\subsubsection{Eigenmodes in Medium $i$}
In any medium, the propagation of an electromagnetic wave is described by exactly four eigenmodes $j=1,2,3,4$ with different $z$-components \qij~of the wavevector. These four \qij~can be obtained for each medium $i$ individually by solving the eigenvalue problem of a characteristic matrix \Del~(Eq.~11 in Ref.~\cite{Passler2017a}), as has been derived originally by Berreman \cite{Berreman1972}. However, for media with highly dispersive permittivities, the four obtained eigenvalues \qij~and their related eigenmodes can switch their order as a function of frequency, and thus have to be identified in an unambiguous manner. Following \etal{Li} \cite{Li1988}, the modes are separated into forward and backward propagating waves according to the sign of \qij~(Eq.~12 in Ref.~\cite{Passler2017a}). We assign the forward propagating (transmitted) waves to be described by $q_{i1}$ and $q_{i2}$, and the backward propagating (reflected) waves by $q_{i3}$ and $q_{i4}$. Furthermore, each pair is sorted by the polarization of the corresponding mode, utilizing the electric fields given by the eigenvectors $\Psi_{ij}$~(Eq.~13 in Ref.~\cite{Passler2017a}). In non-birefringent media, the two modes are separated into p-polarized ($q_{i1}$ and $q_{i3}$) and s-polarized ($q_{i2}$ and $q_{i4}$) waves by analyzing the $x$-component of their electric fields. For birefringent media, on the other hand, the sorting is realized by analyzing the $x$-component of the Poynting vector $\Svec_{ij} = \Evec_{ij} \times \Hvec_{ij}$, and the modes are separated into ordinary ($q_{i1}$ and $q_{i3}$) and extraordinary ($q_{i2}$ and $q_{i4}$) waves \cite{Passler2017a}.

%\subsubsection{\textbf{Transfer Matrix with Treatment of Singularities}}
In the case of non-birefringent media, the four solutions \qij~become degenerate, leading to singularities in the formalisms of previous works \cite{Berreman1972,Yeh1979,Lin-Chung1984}. To resolve this problem, we follow the solution presented by \etal{Xu} \cite{Xu2000}. Using the appropriately sorted $q_{ij}$, obtained as described above, the eigenvectors $\vec{\gamma}_{ij}$ of the four eigenmodes in each layer $i$ are:
\begin{align}
\vec{\gamma}_{ij}=\colvec{\gamma_{ij1} \\ \gamma_{ij2} \\ \gamma_{ij3}},
\end{align}
with the values of $\gamma_{ijk}$ given by \etal{Xu} \cite{Xu2000} (Eq.~20 in Ref.~\cite{Passler2017a}), and $k=1,2,3$ being the $x$, $y$, and $z$ components of $\vec{\gamma}_{ij}$. Furthermore, $\vec{\gamma}_{ij}$ has to be normalized:
\begin{align}
\vec{\hat{\gamma}}_{ij}=\frac{\vec{\gamma}_{ij}}{\abs{\vec{\gamma}_{ij}}}.
\label{eq:eigenmodes}
\end{align}
We note that this normalization is essential to ensure a correct calculation of the cross-polarization components of the transfer matrix. The normalized electric field eigenvectors $\vec{\hat{\gamma}}_{ij}$, being free from singularities, replace the eigenvectors $\Psi_{ij}$ for all further calculations in the formalism. 

At each interface, the boundary conditions for electric and magnetic fields allow to connect the fields of the two adjacent layers $i-1$ and $i$. Formulated for all four modes simultaneously, the boundary conditions are:
\begin{align}
\AMa{i-1} \Evec_{i-1}=\AMa{i} \Evec_i,
\label{eq:A_boundaryShort}
\end{align}
where $\AMa{i}$ is a $4\times4$ matrix calculated from the eigenvectors $\gamma_{ijk}$ \cite{Xu2000} (Eq.~22 in Ref.~\cite{Passler2017a}), and $\Evec_{i}$ is a dimensionless 4-component electric field vector containing the amplitudes of the resulting electric fields of all four modes. In the following, we refer to $\Evec_{i}$ as the amplitude vector, and its components are sorted as follows:
\begin{align}
\label{eq:E_xu_sorting}
\Evec\equiv\colvec{E_{\Rightarrow}^{p/o} \\ E_{\Rightarrow}^{s/e} \\ E_{\Leftarrow}^{p/o} \\E_{\Leftarrow}^{s/e}},
\end{align}
where $\Rightarrow$ ($\Leftarrow$) stands for the forward propagating, transmitted (backward propagating, reflected) modes, and p, s refers to the p- and s-polarized modes in non-birefringent media, while $o$, $e$ indicates the ordinary and extraordinary modes in birefringent media.
By multiplying ${\AMa{i-1}}^{-1}$ on both sides of Eq.~\ref{eq:A_boundaryShort}, we find the implicit definition of the interface matrix \LMa{i}, which projects the amplitude vector in medium $i$ onto the amplitude vector in medium $i-1$:
\begin{align}
\Evec_{i-1}={\AMa{i-1}}^{-1}\AMa{i}\Evec_{i}\equiv \LMa{i}\Evec_{i}.
\end{align}
For the transition between two birefringent or between two non-birefringent media, the projection of a wave of one particular polarization in layer $i$ only yields a finite amplitude in layer $i-1$ of the mode of same polarization, i.e. $s/e \leftrightarrow s/e$, and $p/o \leftrightarrow p/o$. For the transition between a birefringent and a non-birefringent medium, on the other hand, the interface matrix \LMa{i} projects a mode of one particular polarization in layer $i$ onto both polarization states in layer $i-1$. This cross-polarized projection occurs because in birefringent media, the in-plane directions of the ordinary and extraordinary eigenmodes are rotated ($\neq n \pi$, $n \in \mathbb{N}_0 $) with respect to the directions of the p- and s-polarized eigenmodes in the non-birefringent medium. 

The propagation of all four eigenmodes through layer $i$ is described by the propagation matrix \PMa{i} \cite{Yeh1979}:
\begin{align}
\PMa{i}\!=\!\fourfourMatrix
{e^{-i\frac{\omega}{c}q_{i1}d_i} & 0 & 0 & 0}
{0 & e^{-i\frac{\omega}{c}q_{i2}d_i} & 0 & 0}
{0 & 0 & e^{-i\frac{\omega}{c}q_{i3}d_i} & 0}
{0 & 0 & 0 & e^{-i\frac{\omega}{c}q_{i4}d_i}}\!,
\end{align}
where the rotation of polarization in birefringent media arises due to a phase difference that is accumulated during propagation through the medium, because of different propagation speeds of the ordinary and extraordinary modes ($q_{i1} \neq q_{i2}$ and $q_{i3} \neq q_{i4}$).

The transfer matrix \TMa{i}~of a single layer $i$ is defined as:
\begin{align}
\TMa{i}=\AMa{i} \PMa{i} {\AMa{i}}^{-1},
\end{align}
and the full transfer matrix \FullTMa{}~of all $N$ layers is
\begin{align}
\FullTMa{}= {\AMa{0}}^{-1} \left( \prod_{i=1}^{N} \TMa{i} \right) \AMa{N+1},
\end{align}
where ${\AMa{0}}^{-1}$ ($\AMa{N+1}$) ensures the correct mode projection between the multilayer system and the incident medium (substrate).

\subsection{Reflection and Transmission Coefficients}
\label{sec:rt_coefficients}

The full transfer matrix \FullTMa{}~projects the amplitude vector in the substrate $\Evec^+_{N+1}$ onto the amplitude vector in the incident medium $\Evec^-_0$: 
\begin{align}
\Evec^-_0 = \FullTMa{} \Evec^+_{N+1},
\end{align}
where $\Evec^-_{i-1}$ and $\Evec^+_i$ denote the fields on both sides of the interface between layer $i-1$ and $i$, respectively. Following the equations presented by Yeh \cite{Yeh1979}, the transmission ($t$) and reflection ($r$) coefficients for incident p- or s-polarization can be calculated in terms of the matrix elements of \FullTMa{} as follows:
\begin{align}
r_{pp}&=\frac{\FullTMaEl{31}\FullTMaEl{22}-\FullTMaEl{32}\FullTMaEl{21}}{\FullTMaEl{11}\FullTMaEl{22}-\FullTMaEl{12}\FullTMaEl{21}} & 
t_{p(p/o)}&=\frac{\FullTMaEl{22}}{\FullTMaEl{11}\FullTMaEl{22}-\FullTMaEl{12}\FullTMaEl{21}} 
\label{eq:reflectionCoefficient_pp}
\\
r_{ss}&=\frac{\FullTMaEl{11}\FullTMaEl{42}-\FullTMaEl{41}\FullTMaEl{12}}{\FullTMaEl{11}\FullTMaEl{22}-\FullTMaEl{12}\FullTMaEl{21}} &
t_{s(s/e)}&=\frac{\FullTMaEl{11}}{\FullTMaEl{11}\FullTMaEl{22}-\FullTMaEl{12}\FullTMaEl{21}} \\
r_{ps}&=\frac{\FullTMaEl{41}\FullTMaEl{22}-\FullTMaEl{42}\FullTMaEl{21}}{\FullTMaEl{11}\FullTMaEl{22}-\FullTMaEl{12}\FullTMaEl{21}} &
t_{p(s/e)}&=\frac{-\FullTMaEl{21}}{\FullTMaEl{11}\FullTMaEl{22}-\FullTMaEl{12}\FullTMaEl{21}} \\
r_{sp}&=\frac{\FullTMaEl{11}\FullTMaEl{32}-\FullTMaEl{31}\FullTMaEl{12}}{\FullTMaEl{11}\FullTMaEl{22}-\FullTMaEl{12}\FullTMaEl{21}} &
t_{s(p/o)}&=\frac{-\FullTMaEl{12}}{\FullTMaEl{11}\FullTMaEl{22}-\FullTMaEl{12}\FullTMaEl{21}},
\label{eq:reflectionCoefficient_sp}
\end{align}
where the subscripts refer to the incoming and outgoing polarization state, respectively. The transmission coefficients describe the transmitted electric field amplitude into p- and s-polarized states in the case of a non-birefringent substrate, and into the ordinary and extraordinary eigenstates in the case of a birefringent substrate. We note that the indices of \FullTMa{} differ from the equations reported by Yeh \cite{Yeh1979} in order to account for a different order of the eigenmodes in the amplitude vector (Eq.~\ref{eq:E_xu_sorting}).

%\begin{figure*}
%%\includegraphics[width = 1\columnwidth]{fig1}
%\includegraphics[width = .9\textwidth]{figures/fig1.pdf}
%\caption{(a) By means of the transfer-matrix formalism, the \Evec-field distribution can be calculated at any point in the multilayer system. Starting from $\Evec_{N+1}^+$, subsequent multiplication of the interface matrices \LMa{i} and propagation matrices \PMa{i} allows to propagate the wave back to the incident medium, and into the substrate. For the case of evanescent waves, the \Evec-field is exemplary sketched in green. (b) In the Otto geometry (not to scale), a highly refractive prism provides the necessary in-plane momentum for resonant coupling to  SPhPs, propagating along the interfaces of the multilayer system. Here, the SiC / GaN / SiC heterostructure is sketched, discussed in Sec.~3\ref{sec:sim.C}.}
%\label{fig:multilayerStructure}
%\end{figure*} 

\subsection{Electric Field Distribution}
\label{sec:Efield_distribution}

Employing the interface and propagation matrices \LMa{i}~and \PMa{i}, respectively, an amplitude vector can be projected to any $z$-point in the multilayer system (please note the published erratum that corrects the calculation of the electric field distribution in our original work \cite{Passler2019b}). However, due to the rotation of polarization in birefringent media and thus the mixing of polarization states, in general, the cases of incident p- and s-polarization have to be treated seperately. As a starting point, the transmission coefficients can be utilized to formulate the amplitude vector $\Evec_{N+1}^+$ for either p- or s-polarized incident light in the substrate at the interface with layer $N$ as follows:
\begin{align}
\begin{split}
\left( \Evec_{N+1}^+ \right)_{p \text{ in}} &=\colvec{E_{\Rightarrow}^{p/o} \\ E_{\Rightarrow}^{s/e} \\ E_{\Leftarrow}^{p/o} \\E_{\Leftarrow}^{s/e}}=\colvec{t_{p(p/o)} \\ t_{p(s/e)} \\ 0 \\ 0} 
\\
\left( \Evec_{N+1}^+ \right)_{s \text{ in}} &=\colvec{E_{\Rightarrow}^{p/o} \\ E_{\Rightarrow}^{s/e} \\ E_{\Leftarrow}^{p/o} \\E_{\Leftarrow}^{s/e}}=\colvec{t_{s(p/o)} \\ t_{s(s/e)} \\ 0 \\ 0},
\end{split}
\label{eq:Estart}
\end{align}
where the reflected ($\Leftarrow$) components are set to zero, since no light source is assumed to be on the substrate side of the multilayer system. Furthermore, in order to obtain the electric field amplitudes as a function of $z$, the propagation through layer $i$ is calculated by means of the propagation matrix \PMa{i}:
\begin{align}
\begin{split}
\Evec_{i}(z)&=\PMa{i}(z) \Evec^-_{i} \\
&=\!\fourfourMatrix
{\!e^{-i\frac{\omega}{c}q_{i1}z}\! & 0 & 0 & 0}
{0 & \!e^{-i\frac{\omega}{c}q_{i2}z}\! & 0 & 0}
{0 & 0 & \!e^{-i\frac{\omega}{c}q_{i3}z}\! & 0}
{0 & 0 & 0 & \!e^{-i\frac{\omega}{c}q_{i4}z}\!}\!\Evec^-_{i},
\end{split}
\end{align}
with $0<z<d_i$ being the relative $z$-position in layer $i$. Starting from $\Evec_{N+1}^+$, the interface matrices \LMa{i}~and propagation matrices \PMa{i}~subsequently propagate the amplitude vector towards the incident medium. In the reverse direction, the inverse propagation matrix ${\PMa{N+1}}^{-1}$ allows to calculate the \Evec-fields in the substrate. As a result, the four mode amplitudes $E_{\Rightarrow}^{p/o}$, $E_{\Rightarrow}^{s/e}$, $E_{\Leftarrow}^{p/o}$, and $E_{\Leftarrow}^{s/e}$ are obtained as a function of $z$ within each layer. 

In order to obtain the three-components $\E_x$, $\E_y$, and $\E_z$ of the electric field for each of the four modes $j$, the four mode amplitudes have to be multiplied with their respective eigenmode vector $\vec{\hat{\gamma}}_{ij}$ (Eq. \ref{eq:eigenmodes}). This yields for the electric fields $\vec{\E}$ of the four modes for each layer $i$, as a function of $z$, and for either p- or s-polarized incident light:
\begin{align}
\begin{split}
\left( \vec{\E}_{\Rightarrow}^{p/o} \right)_{p/s \text{ in}} &= \left( E_{\Rightarrow}^{p/o} \right)_{p/s \text{ in}} \colvec{ \hat{\gamma}_{i11} \\ \hat{\gamma}_{i12}  \\  \hat{\gamma}_{i13} }, \quad j=1 \\
\left( \vec{\E}_{\Rightarrow}^{s/e} \right)_{p/s \text{ in}} &= \left( E_{\Rightarrow}^{s/e} \right)_{p/s \text{ in}} \colvec{ \hat{\gamma}_{i21} \\ \hat{\gamma}_{i22}  \\  \hat{\gamma}_{i23} }, \quad j=2 \\
\left( \vec{\E}_{\Leftarrow}^{p/o} \right)_{p/s \text{ in}} &= \left( E_{\Leftarrow}^{p/o} \right)_{p/s \text{ in}} \colvec{ \hat{\gamma}_{i31} \\ \hat{\gamma}_{i32}  \\  \hat{\gamma}_{i33} }, \quad j=3 \\
\left( \vec{\E}_{\Leftarrow}^{s/e} \right)_{p/s \text{ in}} &= \left( E_{\Leftarrow}^{s/e} \right)_{p/s \text{ in}} \colvec{ \hat{\gamma}_{i41} \\ \hat{\gamma}_{i42}  \\  \hat{\gamma}_{i43} }, \quad j=4,
\label{eq:EfieldVectors}
\end{split}
\end{align}
where we have omitted the index $i$ and the $z$ dependence for the sake of readability. The full electric field $\vec{\E}_i(z)$ for either p- or s-polarized incident light in layer $i$ at point $z$ is given by the sum of all four electric field vectors: 
\begin{align}
\vec{\E}_i(z) = \vec{\E}_{\Rightarrow,i}^{p/o}(z) + \vec{\E}_{\Rightarrow,i}^{s/e}(z) + \vec{\E}_{\Leftarrow,i}^{p/o}(z) + \vec{\E}_{\Leftarrow,i}^{s/e}(z). 
\label{eq:FullE}
\end{align}
The in-plane components of the sum, $\E_x$ and $\E_y$, are continuous throughout the entire multilayer structure, as it is required by Maxwells boundary conditions.

\section{Layer-Resolved Transmittance and Absorption}
\label{sec:layer_res_trans_abs}
The total reflectance \R~of the multilayer system for a given ingoing and outgoing polarization $a$ and $b$, respectively, can be readily calculated from the corresponding reflection coefficient (Eqs. \ref{eq:reflectionCoefficient_pp}-\ref{eq:reflectionCoefficient_sp}):
\begin{align}
\R^{ab} = \abs{r_{ab}}^2, \qquad a,b = p,s.
\label{eq:Rfromr}
\end{align}
The transmittance \T, which is the transmitted power into the substrate, on the other hand, in general is not given by the electric field intensity $\T \neq \abs{t}^2$ (except for the special case if the substrate is vacuum, $\epsi = 1$). Instead, \T~can be calculated from the time-averaged Poynting vector $\vec{\SSS}$ \cite{Yariv1984,ALBERDI2002,Weber2014}, which describes the direction and magnitude of the energy flux of an electromagnetic wave at any point $z$ in the structure:
\begin{align}
\vec{\SSS}_i(z) = \frac{1}{2} \text{Re} \left[ \vec{\E}_i(z) \times \vec{\HH}_i^*(z) \right].
\label{eq:Svec}
\end{align}
The full electric field $\vec{\E}$ (for incident polarization $a$) as a function of $z$ in each layer $i$ was calculated in the previous section (Eq. \ref{eq:FullE}), and the full magnetic field $\vec{\HH}$~is obtained as follows using Maxwells equations:
\begin{align}
\begin{split}
\vec{\HH}_i(z) = \frac{1}{\omega \mu_i} \Bigl( &\kvec_{i1} \times \vec{\E}_{\Rightarrow,i}^{p/o}(z) + \kvec_{i2} \times \vec{\E}_{\Rightarrow,i}^{s/e}(z) \\
                                                         +&\kvec_{i3} \times \vec{\E}_{\Leftarrow,i}^{p/o}(z) + \kvec_{i4} \times \vec{\E}_{\Leftarrow,i}^{s/e}(z) \Bigr),
%\frac{1}{\omega \mu_i} \sum_{j=1}^{4} \kvec_{ij} \times \vec{\E}_{ij}(z),
\end{split}
\label{eq:FullH}
\end{align}
where $\kvec_{ij}$ are the wavevectors in layer $i$ of the four modes $j$, see Eq. \ref{eq:kvec}. Because $\vec{\E}$ and $\vec{\HH}$ are known in each layer $i$ and as a function of $z$ from the transfer matrix formalism, the Poynting vector can be evaluated likewise yielding $\vec{\SSS}_{i}(z)$, which will be used in the following to calculate the transmittance and absorption at any point $z$ in the multilayer system.

It is important to note that while $\vec{\E}$ and $\vec{\HH}$ can be calulcated for each of the four modes $j$ individually, this mode separation in general -- specifically, in the case of birefringent media -- is not possible for the Poynting vector $\vec{\SSS}$. In other words, in birefringent media, the sum of the Poynting vectors of the four modes is not equal to the Poynting vector calculated from the total fields $\vec{\E}$ (Eq. \ref{eq:FullE}) and $\vec{\HH}$ (Eq. \ref{eq:FullH}). The difference arises because in birefringent media, $\vec{\E} \not\perp \vec{\HH}$. Therefore, the cross products $\vec{\E} \times \vec{\HH}$ between different modes $j$ are no longer zero. For the correct calculation of the Poynting vector in birefringent media, it is thus necessary to calculate the cross product of the total fields $\vec{\E}$ and $\vec{\HH}$, as shown in Eq. \ref{eq:Svec}. Interestingly, this means that the energy flux in birefringent media cannot be split up into the ordinary and extraordinary eigenmodes, but has to be considered as a single quantity. In the following, we therefore discuss the transmittance and absorption for s- or p-polarized incident light without differentiating between the eigenmodes anymore. 

An exception is the incident medium, which is set to be isotropic. Here, the Poynting vector can be calculated for each mode individually, and for the purpose of normalizing the transmitted power, we calculate the Poynting vector of the incident light $\vec{\SSS}_{\text{inc}}$ (in layer $i=0$ at position $z=0$) for either p- or s-polarization as follows:
\begin{align}
\begin{split}
\vec{\SSS}_{\text{inc}}^{p} &= \frac{1}{2} \text{Re} \left[ \vec{\E}_{\Rightarrow,0}^{p}(0) \times \left( \kvec_{01} \times \vec{\E}_{\Rightarrow,0}^{p}(0)\right)^* \right] \\
\vec{\SSS}_{\text{inc}}^{s} &= \frac{1}{2} \text{Re} \left[ \vec{\E}_{\Rightarrow,0}^{s}(0) \times \left( \kvec_{02} \times \vec{\E}_{\Rightarrow,0}^{s}(0)\right)^* \right].
\end{split}
\label{eq:Sinc}
\end{align}
In a stratified multilayer system, the transmitted energy is given by the $z$-component of the Poynting vector. Thus, we note that alternatively to Eq. \ref{eq:Rfromr}, the reflectance \R~can be calculated from the Poynting vector:
\begin{align}
\R^{ab} = \frac{-\SSS_{\text{refl},z}^{b}}{\SSS_{\text{inc},z}^{a}},
\end{align}
where the minus sign accounts for the negative $z$-direction of the reflected light. $\SSS_{\text{refl},z}^{b}$ is the $z$-component of the Poynting vector of the reflected light of polarization $b=p,s$ (in layer $i=0$ at position $z=0$) given by:
\begin{align}
\begin{split}
\vec{\SSS}_{\text{refl}}^{p} &= \frac{1}{2} \text{Re} \left[ \vec{\E}_{\Leftarrow,0}^{p}(0) \times \left( \kvec_{03} \times \vec{\E}_{\Leftarrow,0}^{p}(0)\right)^* \right] \\
\vec{\SSS}_{\text{refl}}^{s} &= \frac{1}{2} \text{Re} \left[ \vec{\E}_{\Leftarrow,0}^{s}(0) \times \left( \kvec_{04} \times \vec{\E}_{\Leftarrow,0}^{s}(0)\right)^* \right].
\end{split}
\label{eq:Srefl}
\end{align}
As discussed above, for anisotropic media, a separation of the energy flow into the different eigenmodes of polarization $b$ is not generally possible. Therefore, all following equations calculate the total transmittance or absorption for the respective incident polarization $a$.

\begin{figure*}[!ht]
\includegraphics[width = \textwidth]{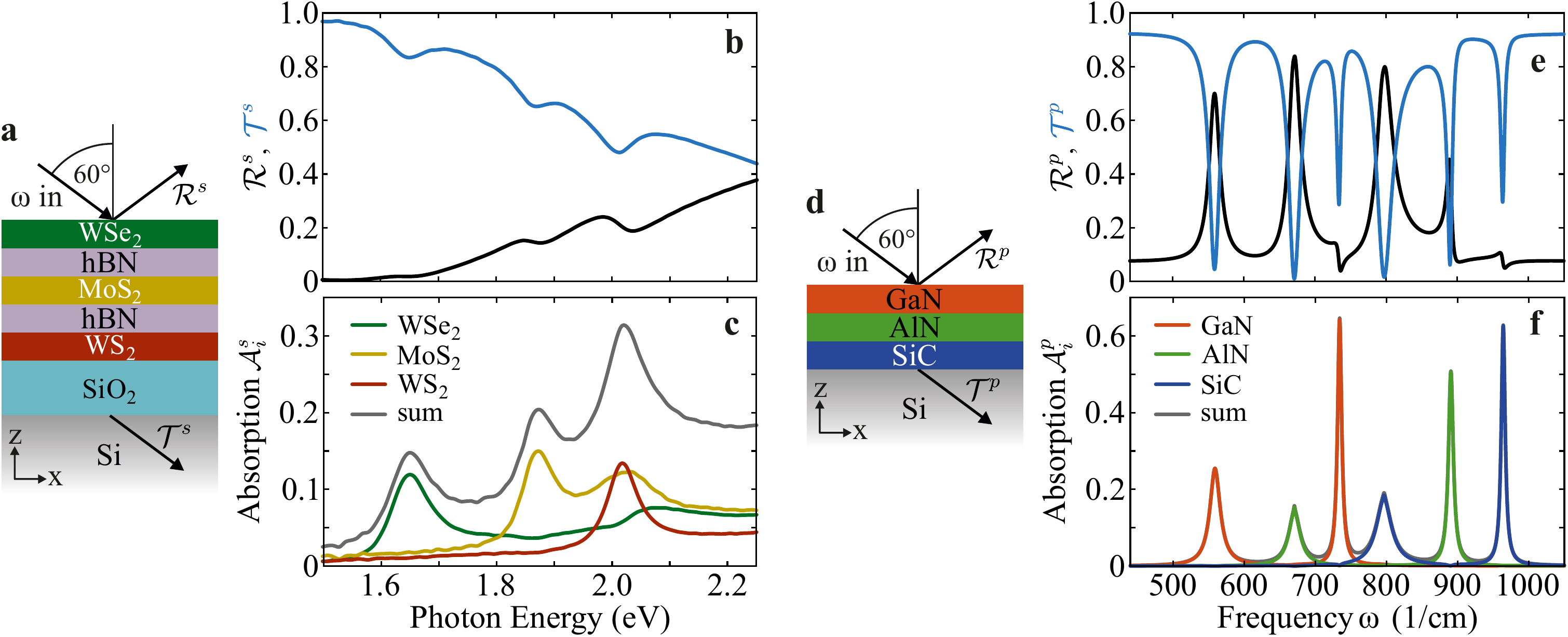}
\caption{\textbf{Layer-resolved absorption calculations of two simple test structures.} \textbf{a} Sketch of the TMDC heterostructure comprising monolayers of \WSe, \MoS, and \WS~\cite{Jung2019} separated by hBN monolayers on a \nmetr{140} thick \SiO~film on a Si substrate. \textbf{b} Reflectance (black line) and transmittance (blue line) spectra of the system at visible wavelengths and at an incident angle of \dg{60} for s-polarized incident light. \textbf{c} Layer-resolved absorption spectra of the TMDC monolayers (colored lines), revealing their respective contribution to the overall absorption spectrum (gray line). \textbf{d} Sketch of the polar dielectric heterostructure comprising \nmetr{100} thin GaN, AlN, and SiC films on a Si substrate. \textbf{e} Reflectance and transmittance spectra and \textbf{f} layer-resolved absorption spectra of the second system, enabling identification of the absorbing layer for each of the six absorption lines.}
\label{fig:testCases}
\end{figure*}  

The transmittance \T~into the substrate $i=N+1$ at the interface with layer $N$ for incident light of polarization $a$ is given by:
\begin{align}
\T^{a} = \frac{\SSS_{N+1,z}^{a}(D)}{\SSS_{\text{inc},z}^{a}},
\end{align}
where $D=\sum_{i=1}^{N} d_i$ is the thickness of the multilayer system. The full $z$-dependence of the transmittance can be evaluated by using the $z$-component of the Poynting vector $\SSS_{i,z}(z)$ at a certain $z$-position in layer $i$:
\begin{align}
\T_i^{a}(z) = \frac{\SSS_{i,z}^{a}(z)}{\SSS_{\text{inc},z}^{a}}.
\end{align}
With this, the absorption \A~of the entire multilayer system, that is up to the last interface between layer $N$ and the substrate, is given by
\begin{align}
\A^{a} = 1-\R^{a}-\T^{a},
\label{eq:Atotal}
\end{align}
and the $z$-resolved absorption $\A_i(z)$ in each layer $i$ is
\begin{align}
\A_i^{a}(z) = 1-\R^{a}-\T_i^{a}(z),
\label{eq:Aiz}
\end{align}
where $\R^{a} = \R^{ap}+\R^{as}$ is the total reflectance. Note that Eq. \ref{eq:Aiz} describes the total absorption starting from $z=0$ at the first interface up to the specified position $z$ in layer $i$. The layer-$i$-resolved absorption, on the other hand, is given by:
\begin{align}
\begin{split}
\A_i^a &= \T_i^a(d_{1..i-1}) - \T_i^a(d_{1..i-1} + d_i) \\ 
		&=  \A_i^a(d_{1..i-1} + d_i) - \A_i^a(d_{1..i-1}),
\end{split}
\label{eq:Ai}
\end{align}
where $d_{1..i-1} = \sum_{i=1}^{i-1} d_i$ is the thickness of all layers through which the incident light has propagated before reaching the layer $i$.

Before we study three nanophotonic device structures in the following section, we calculate the layer-resolved absorption for two simple test structures in Fig.~\ref{fig:testCases}. The first is a typical TMDC heterostructure, comprising monolayers of tungsten diselenide (\WSe), \MoS, and tungsten disulfide (\WS)~sandwiched between hBN monolayers (sketched in Fig.~\ref{fig:testCases}a), where each TMDC monolayer features individual exciton absorption lines. In Fig.~\ref{fig:testCases}b, the reflection and transmittance of the entire structure is plotted, and Fig.~\ref{fig:testCases}c shows the absorption spectra of each TMDC monolayer. The total absorption (gray line in Fig.~\ref{fig:testCases}c) obtained from the reflectance and transmittance spectra (Eq.~\ref{eq:Atotal}) exhibits three indistinguishable absorption features, whereas the layer-resolved absorption calculations unravel the absorption spectrum, allowing to identify the contribution of each TMDC monolayer.

In Fig.~\ref{fig:testCases}d-f, we show the absorption in a polar dielectric heterostructure of SiC, AlN, and GaN thin films on a Si substrate probed at infrared (IR) frequencies. Polar dielectric crystals feature an IR-active transverse optical (TO) phonon mode, where light is predominantly absorbed, whereas the longitudinal optical (LO) phonon mode is not IR-active and thus featureless in a bulk crystal. Thin films, on the other hand, support the so called Berreman mode in proximity to the LO frequency, leading to a strong absorption feature at oblique incidence \cite{Berreman1963,Vassant2012,Passler2019,Dunkelberger2020}. Thus, the reflectance and transmittance spectra (Fig.~\ref{fig:testCases}e) of the polar dielectric heterostructure are of complicated shape, exhibiting six different features arising from the three different polar crystal thin films. The layer-resolved absorption calculations split these features into three spectra with two absorption peaks each (Fig.~\ref{fig:testCases}f), allowing to identify the respective polar crystal thin film that leads to the absorption at its respective TO and LO frequencies.

\section{Simulations of Nanophotonic Devices}
\label{sec:simulations}
The transfer matrix formalism and the calculation of the layer-resolved absorption and transmittance presented in the previous section can be applied for any wavelength and any number of layers, consisting of birefringent or non-birefringent media described by an arbitrary permittivity tensor $\epsiTens_i$. As case studies, in this section we describe three selected nanophotonic device structures. The first example discusses the hyperbolic phonon polaritons arising in a \MoO~/ AlN / SiC system excited evanescently at far-IR wavelengths, highlighting the potential of the formalism for nanophotonic studies in stratified media. The second example describes the freespace response in the visible of a van der Waals heterostructure of monolayers of \MoS~embedded in a hBN matrix, featuring layer-selective absorption of the \MoS~excitons. Finally, the third example calculates the mid-IR absorption of strongly coupled modes formed from intersuband plasmons in multi-quantum wells embedded in an optical cavity.

\begin{figure*}[!ht]
\includegraphics[width = \textwidth]{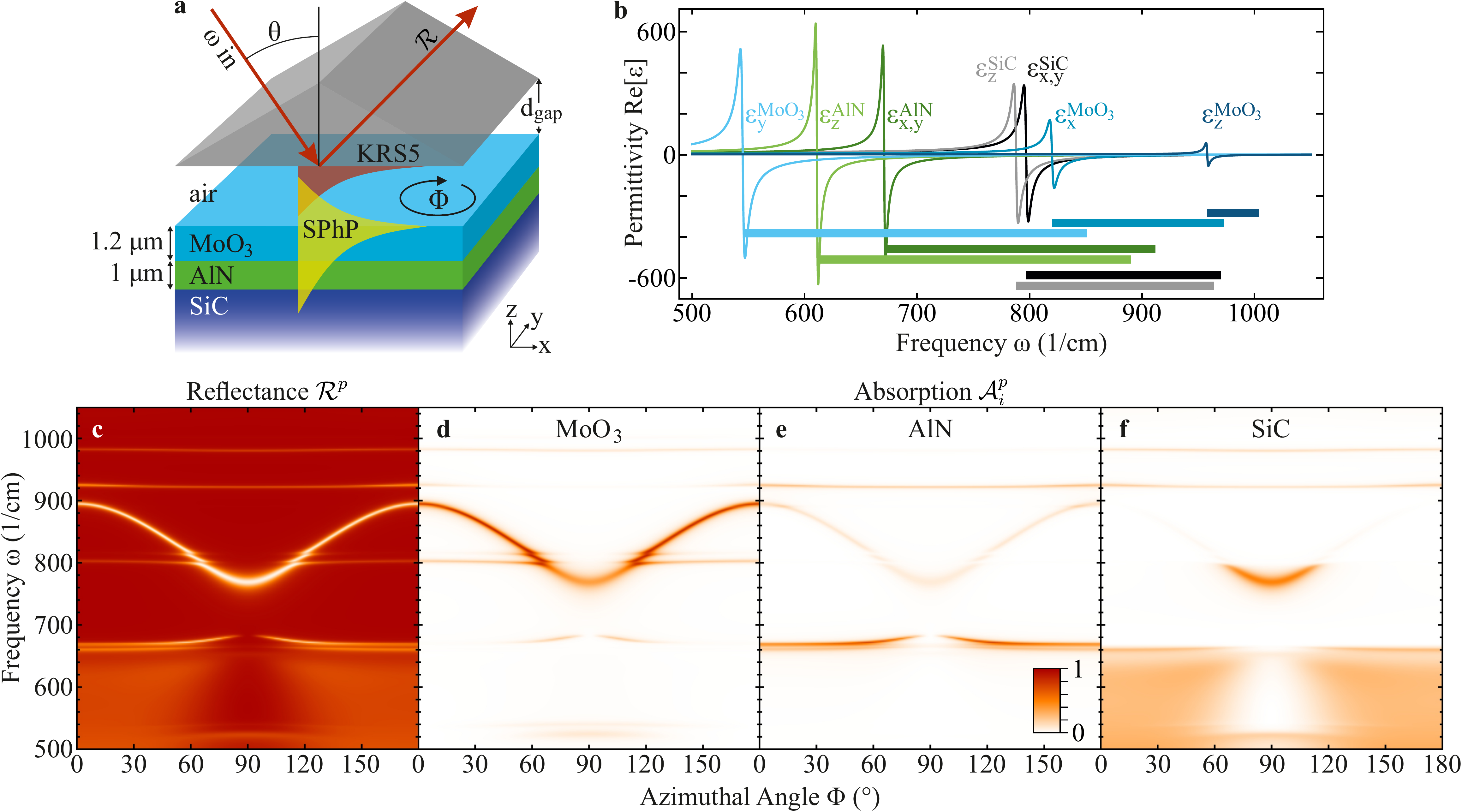}
\caption{\textbf{Azimuthal dependence of the layer-resolved absorption in a \MoO/AlN/SiC heterostructure.} \textbf{a} Prism coupling setup in the Otto geometry for the evanescent excitation of phonon polaritons in a \nmetr{1200} \MoO~/ \nmetr{1000} AlN / SiC substrate multilayer sample with p-polarized incident light. The incident angle is fixed to $\theta=\dg{28}$ and the air gap to $d_{\text{gap}}=\mumetr{8}$. \textbf{b} Real value of the principle relative permittivities $\epsi_x$, $\epsi_y$, and $\epsi_z$ of uniaxial SiC and AlN and biaxial \MoO. Their respective reststrahlen bands located between the corresponding TO and LO phonon frequencies are indicated by the horizontal color bars. \textbf{c} Reflectance $\R^p$ of the entire system as a function of incident frequency $\omega$ and azimuthal angle of the sample $\Phi$. \textbf{d-f} Absorptions $\A_i^p$ (Eq. \ref{eq:Ai}) of the \MoO, AlN, and SiC layers as a function of $\omega$ and $\Phi$.}
\label{fig:MoO3}
\end{figure*}  

\subsection{Hyperbolic Phonon Polaritons in \MoO/AlN/SiC}
\label{sec:MoO3}

\begin{figure*}[!ht]
\includegraphics[width = \textwidth]{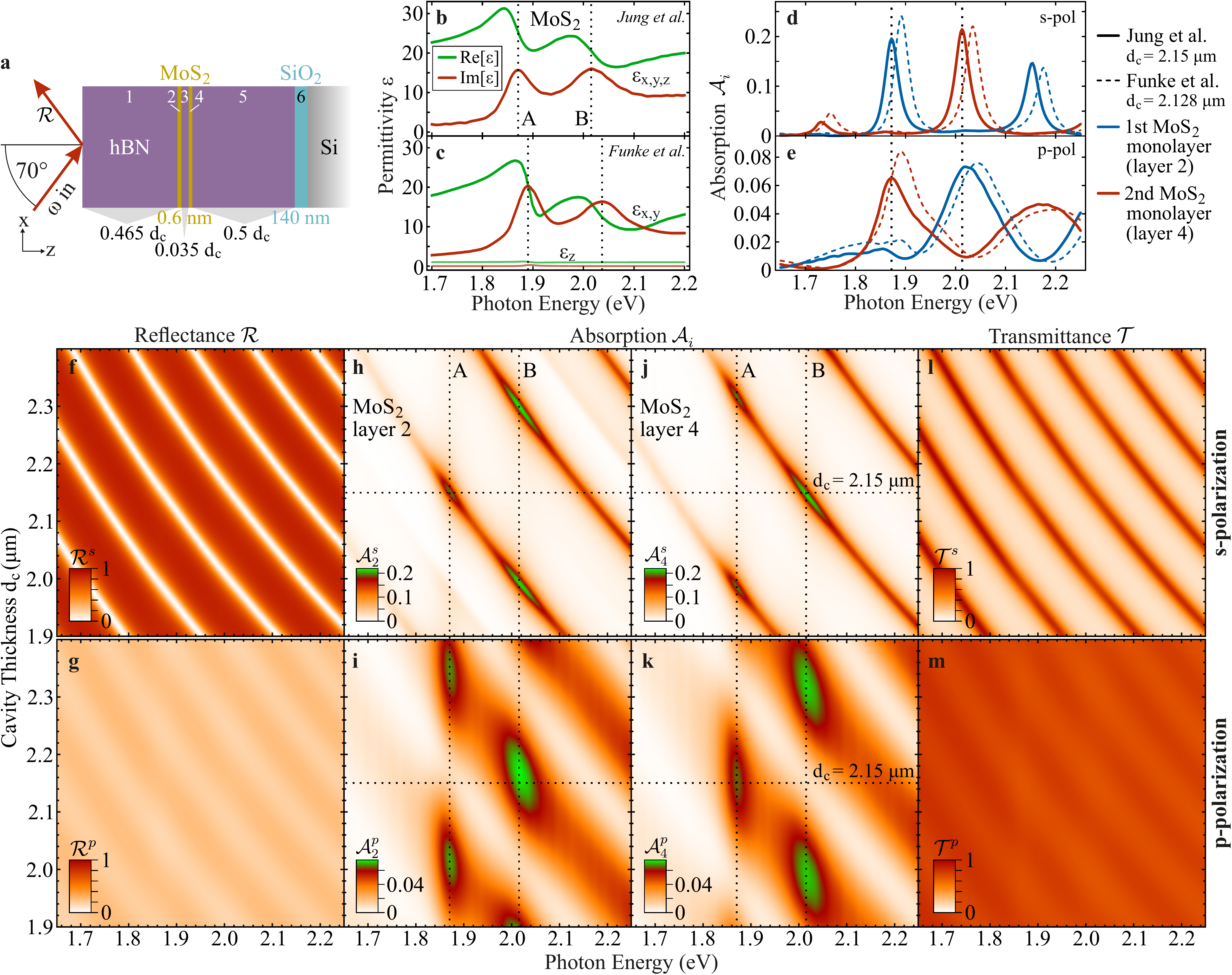}
\caption{\textbf{Cavity-enhanced exciton absorption in \MoS~monolayers.} \textbf{a} Structure of the Fabry-P\'{e}rot cavity. A \nmetr{140} thin \SiO~film on the Si substrate serves as a back-reflector, while the cavity with variable thickness $d_c=1.9-\mumetr{2.4}$ is made of three layers of hBN of relative thicknesses $0.465$ $d_c$, $0.035$ $d_c$, and $0.5$ $d_c$, separated by two \nmetr{0.6} thin \MoS~monolayers. \textbf{b} Real and imaginary values of the isotropic relative permittivity of \MoS~\cite{Jung2019}. All calculations are based on this permittivity, except the dashed lines in subfigures d and e. \textbf{c} Anisotropic relative permittivity of \MoS~\cite{Funke2016}. \textbf{d} Absorption spectra of the two \MoS~monolayers for s-polarized incident light, for a cavity thickness of $d_c=\mumetr{2.15}$ ($d_c=\mumetr{2.128}$) optimized for layer-selective absorption of the two \MoS~excitons, calculated with the \MoS~permittivities of \etal{Jung} (\etal{Funke}). \textbf{e} Analogous calculations as in subfigure d but for p-polarized incident light. \textbf{f,g} Reflectance $\R^s$ and $\R^p$ for s- and p-polarized incident light, respectively. \textbf{h,i} and \textbf{j,k} Absorption $\A_i^s$ and $\A_i^p$ of the two \MoS~monolayers, respectively. \textbf{l,m} Transmitted power $\T^s$ and $\T^p$ into the Si substrate, respectively. The maps of \R, \A, and \T~are all plotted as a function of the incident photon energy and the cavity thickness $d_c$.}
\label{fig:MoS2}
\end{figure*}  

Polar crystals such as \MoO, AlN, and SiC support surface phonon polaritons (SPhP) at frequencies inside their reststrahlen region between the TO and LO phonon frequencies \cite{Caldwell2015}. On smooth surfaces, SPhPs can be excited via prism coupling in the Otto geometry \cite{Otto1968,Neuner2009} as illustrated in Fig.~\ref{fig:MoO3}a, where the air gap width $d_{\text{gap}}$ governs the excitation efficiency and the incident angle $\theta$ defines the in-plane momentum of the launched SPhP \cite{Passler2017}. The system we investigate here is a multilayer heterostructure comprising an \MoO~and AlN film on a SiC substrate. Employing the presented transfer matrix fomalism, the layer-resolved absorption $\A_i$ of such a structure can be calculated as a function of incident angle $\theta$, incident frequency $\omega$, layer thicknesses $d_i$, and azimuthal angle~$\Phi$ of the sample.

The absorption in the \MoO, AlN, and SiC layers as a function of $\omega$ and $\Phi$ and for fixed $\theta$ and $d_i$ is shown in Fig.~\ref{fig:MoO3}d-f. The reflectance of the entire system is plotted in Fig.~\ref{fig:MoO3}c. As required by energy conservation, the sum of the absorbed power in the three polar crystals equals the attenuated power visible as absorption dips in the reflectance. However, while the reflectance only yields the total absorption, the layer-resolved calculations allow to identify the exact position of a power drain in a multilayer system. 

In particular, the \MoO/AlN/SiC heterostructure features several sharp absorption lines at $660$, $800$, $920$, and \wavenumber{980} that are mostly independent of $\Phi$, and one prominent absorption line that strongly varies with $\Phi$, indicating that the latter depends on in-plane anisotropy ($\epsi_x \neq \epsi_y$) while the former do not. In the multilayer sample, only the \MoO~layer exhibits in-plane anisotropy, while AlN and SiC are c-cut uniaxial crystals (principle relative permittivities $\epsi_x$, $\epsi_y$, and $\epsi_z$ are shown in Fig.~\ref{fig:MoO3}b). Indeed, the $\Phi$-dependent feature is mostly absorbed in the \MoO~layer. This feature is the hyperbolic phonon polariton (hPhP) supported by the air/\MoO~interface arising in the in-plane reststrahlen bands of \MoO. Its tunability in frequency arises from the large in-plane anisotropy of \MoO, which leads to a $\Phi$-dependent effective permittivity sensed by the hPhP upon azimuthal rotation. 

Notably, below the SiC TO frequencies ($\sim\wavenumber{800}$), a significant part of the hPhP leaks into the SiC substrate, while above \w{TO}{SiC}~the mode is mostly confined to the \MoO~layer. The high confinement above \w{TO}{SiC}~occurs because of the negative permittivity in the SiC reststrahlen band, while below \w{TO}{SiC}, the mode can penentrate the substrate. Interestingly, this mode penetration happens across the AlN layer, where only a small part of the mode is absorbed. Since AlN features its reststrahlen bands across the entire frequency range of the hPhP supported by the \MoO, and thus evanescently attenuates all modes, the hPhP appears to tunnel through the AlN layer to be absorbed by the SiC substrate. 

The multilayer system presented here only scratches the surface of various possible material configurations and compositions that can potentially be employed for tailoring surface and interface polariton resonances \cite{Jia2015,Dufferwiel2015,Low2017,Wintz2018,Passler2019a,Ratchford2019,Zhang2019}. In particular the emerging field of volume-confined hyperbolic polaritons \cite{Li2015,Ma2018,Dai2018,Ratchford2019}, enabled by the anisotropic permittivity of the supporting media, holds great potential for future nanophotonic applications, such as subdiffraction imaging and hyperlensing \cite{Dai2015,Ferrari2015}. Providing the full layer-resolved absorption, our algorithm paves the way to predict and study hyperbolic polariton modes in any anisotropic stratified heterostructure.

\subsection{Layer-selective absorption of \MoS~excitons in a Fabry-P\'{e}rot cavity}
\label{sec:MoS2}
TMDC monolayers such as \MoS~feature strong exciton resonances at visible frequencies \cite{Funke2016,Jung2019}. By inserting these monolayers into van der Waals heterostructures forming a Fabry-P\'{e}rot cavity, the light-matter interaction enabled by the TMDC exciton can be strongly enhanced \cite{Schneider2018}. Here, we embed two \MoS~monolayers into a $d_c =1.9-\mumetr{2.4}$ thick hBN cavity with a \SiO~back-reflector on a Si substrate \cite{Benameur2011}, as sketched in Fig.~\ref{fig:MoS2}a. Employing the presented formalism, the absorption $\A_i$ in the two \MoS~monolayers as a function of photon energy and cavity thickness $d_c$ can be calculated.

In the photon energy range of $1.7-2.2$ eV, \MoS~features two excitons (A and B) at $\sim 1.9$ and $\sim 2.1$ eV. These excitons are apparent as resonance peaks in the isotropic relative permittivity plotted in Fig.~\ref{fig:MoS2}b and resulting in peaks in an absorption spectrum. However, the Fabry-P\'{e}rot cavity creates a static modulation of the electric field enhancement with peaks and nodes as a function of $z$ position, and thus the resulting absorption in a \MoS~monolayer not only depends on the photon energy, but also sensitively depends on the $z$ position of the \MoS~monolayer in the cavity. Taking advantage of this field modulation, we place one \MoS~monolayer (layer 4) at the center of the cavity where the cavity modes alternate between node and peak with maximal amplitude, and the other \MoS~monolayer (layer 2) in close proximity where the cavity features a node when there is a peak in the center, and vice versa.

The cavity modes yield the periodic modulation in photon energy and cavity thickness $d_c$ that can be seen in the reflectance (transmittance) maps shown in Fig.~\ref{fig:MoS2}f and g (l and m), for incoming s- and p-polarized light, respectively. The different modulation contrast for $\R^s$ and $\R^p$ ($\T^s$ and $\T^p$) arises from the large incident angle of $\theta = \dg{70}$, which was chosen to optimize the absorption $\A^s$ in the \MoS~monolayers for s-polarized incident light. For smaller incident angles, the differences for s- and p-polarization decrease, but with a reduction in the absorption $\A^s$. 

In Fig.~\ref{fig:MoS2}h,i (j,k) the absorption for s- and p-polarized incident light in the first \MoS~monolayer, $\A_2^{s,p}$ (second \MoS~monolayer, $\A_4^{s,p}$), is shown. Due to the choice of the $z$ positions of the two \MoS~monolayers, each film is sensitive to only every second cavity mode, where layer 2 (first \MoS~monolayer) absorbs those modes that are not absorbed by layer 4 (second \MoS~monolayer). Additionally to the absorption modulation imposed by the cavity, the A and B excitons of \MoS~yield two absorption features at their respective energies, marked by dotted vertical lines in Fig.~\ref{fig:MoS2}h-k. For the optimized case of s-polarized incident light, the \MoS~monolayers reach cavity-enhanced absorption values of up to 20\% at both exciton energies. At a cavity thickness of $d_c = \mumetr{2.15}$ for s-polarization, layer 2 only absorbs at the energy of exciton A, while in layer 4, absorption only occurs at the energy of exciton B. This layer-selective absorption is further illustrated in the absorption spectra (solid lines) for a fixed cavity thickness of $d_c = \mumetr{2.15}$ shown in Fig.~\ref{fig:MoS2}d and e for s- and p-polarized light, respectively. For both polarizations, a high contrast between the two layers at each exciton absorption line is achieved, yielding efficient layer-selectivity. 

Finally, we compare these results obtained for an isotropic permittivity model \cite{Jung2019} with the same calculations performed for an anisotropic model of \MoS~\cite{Funke2016}. In Fig.~\ref{fig:MoS2}c, the in-plane ($\varepsilon_{x,y}$) and out-of-plane ($\varepsilon_{z}$) permittivity values taken from \etal{Funke} \cite{Funke2016} are plotted. While $\varepsilon_{x,y}$ are qualitatively the same as the values from \etal{Jung} \cite{Jung2019} (Fig.~\ref{fig:MoS2}b), $\varepsilon_{z}$ differs strongly, taking the almost constant value of $\varepsilon_{z}=1+0\: i$. Even though the difference in  $\varepsilon_{z}$ is substantial, the absorption spectra shown in Fig.~\ref{fig:MoS2}d,e (dashed lines) are qualitatively identical to the spectra calculated from an isotropic permittivity model (solid lines). This confirms that spectroscopic measurements of TMDC monolayers are mostly insensitive to their out-of-plane permittivity \cite{Jung2019},  with the exception of cases where $\varepsilon_{z}$ features a zero-crossing, the so-called epsilon-near-zero (ENZ) frequency, giving rise to drastic optical responses such as enhanced higher-harmonic generation \cite{Vincenti2011,Capretti2015a,Passler2019}. However, this is not the case for \MoS, and therefore the spectra are almost identical.

In this example, the layer-resolved absorption calculations from our algorithm provide the essential information for simulating the layer-selective exciton absorption and optimizing the system parameters. In the thriving field of 2D nanophotonics featuring TMDC van der Waals heterostructures, where structures are optimized for maximal light harvesting \cite{Fortin1982,Yu2013}, optoelectronic devices \cite{He2013,Ross2014} or nanolasers \cite{Gan2013,Sobhani2014}, such layer-resolved absorption calculations promise to be of essential importance. Due to the generality of the presented algorithm, the light-matter interaction in any 2D heterostructure can be readily investigated, highlighting the broad applicability of our approach.

\begin{figure*}[!ht]
\includegraphics[width = \textwidth]{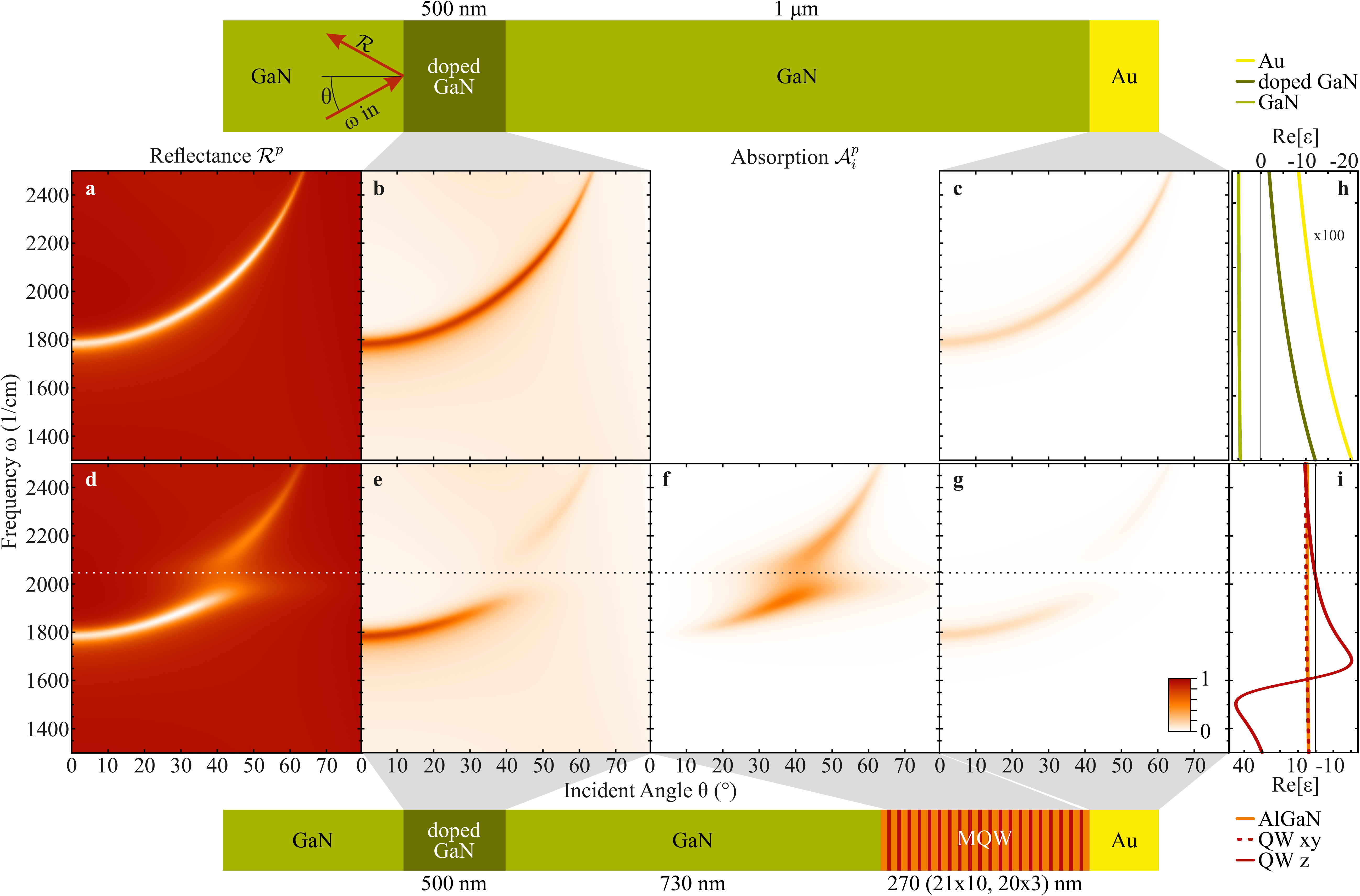}
\captionof{figure}{\textbf{Strong coupling between an ENZ mode in a MQW superlattice and a cavity mode.} 
	A \mumetr{1} thick GaN cavity is formed between a Au mirror and a \nmetr{500} thick doped GaN slab on a GaN substrate (top sketch). 
	The doped GaN slab acts as a low-index mirror. 
	\textbf{a} Simulated reflectance $\R^p$ of the cavity mode showing its dispersion relation as a function of 
	the angle of incidence $\theta$ inside the GaN substrate. 
	\textbf{b,c} Layer-resolved absorption $\A^p$ inside the doped GaN slab 	and in the Au mirror, respectively. 
	\textbf{d} Calculated reflectance $\R^p$ of the \mumetr{1} thick cavity partially filled with a  doped GaN/AlGaN 
	MQW superlattice (see bottom sketch), featuring an electronic excitation at $\sim \wavenumber{2030}$ (white dotted line). 
	The strong light-matter coupling between the cavity mode and QW resonance leads to an avoided crossing and the emergence of two polariton branches. 
	\textbf{e-g} Layer-resolved absorption $\mathcal{A}^{p}$ inside the doped GaN mirror, the superlattice and the Au mirror, respectively. 
	\textbf{h} Real part of the permittivity of the isotropic layers forming the cavity. 
	\textbf{i} Real part of the different components of the permittivity tensor of the doped GaN QW and of the AlGaN barrier, 
	showing the ENZ mode frequency of the QW at  $\sim \wavenumber{2030}$.}
\label{fig:MQW}
\end{figure*}

\subsection{Strong Coupling in a Multi Quantum Well - Cavity System}
\label{sec:MQW}
Doped semiconductor quantum wells (QWs) support transitions between consecutive quantum-confined electronic levels, 
called intraband or inter-subband (ISB) transitions. Contrary to interband transitions, ISB transitions do not only depend 
on the bandgap properties of the semiconductor, but also on the electronic confinement inside the QWs, and thus offer a great 
frequency tunability by changing the width, but also the doping level inside the QW. 
They play a major role in semiconductor optic devices operating in the IR where 
semiconducting materials with a suitable bandgap are lacking, and are the building block of 
quantum well IR photodetectors \cite{Levine1987} and quantum cascade lasers \cite{Faist1994}. 
They also offer a practical platform to study the optical properties of dense confined electron gases \cite{Vasanelli2016}, 
which notably lead to the demonstration of the strong \cite{Liu1997,Dini2003} and ultra-strong 
light-matter coupling regimes \cite{Ciuti2005, Todorov2010}. 
It is remarkable that such fundamental electrodynamical phenomena are directly observable on semiconductor devices \cite{Dupont2003, Jouy2010,Vigneron2019}. 
One peculiar aspect of these ISB transitions is that they only couple to the component of the electric field along 
the confinement direction of the QW structure. Hence, the optical properties of a doped QW can be described by an effective 
permittivity tensor with different in-plane ($\varepsilon_{x,y}$) and out-of-plane ($\varepsilon_{z}$) values, which has been realized by several permittivity models \cite{Ando1982,Zauzny1999,Alpeggiani2014,Pegolotti2014}. The description of light propagation in stratified anisotropic media containing such QWs requires a complex formalism, such as the transfer matrix method the here presented algorithm builds on \cite{Passler2017a}. 

We focus here on an existent experimental configuration to further demonstrate the potential of our formalism for the case of strong light-matter coupling between a cavity mode and a collective intersubband excitation in a multi-quantum well (MQW) superlattice. The system is composed of a GaN cavity formed by a 
\nmetr{500} thick, Si-doped GaN slab and a Au mirror, as sketched on the top of Fig.~\ref{fig:MQW}. 
The doped GaN layer is modeled using the Drude model, and acts as a low-index mirror. 
The two mirrors are separated by a \mumetr{1} thick GaN spacer, forming an empty cavity.  
The system sustains a guided transverse magnetic (TM) mode, where the electric field is confined mostly between the two mirrors and its out-of-plane component is maximal near the Au mirror. This guided mode can be probed in a reﬂectance experiment, as discussed in the following. In order to probe large internal angles of incidence experimentally, the sample has to be prepared in a prism shape, for example by cleaving the incident GaN layer facets. 
The calculated p-polarized reflectance $\R^{p}$ is shown in Fig.~\ref{fig:MQW}a, 
evidencing the dispersion relation of the cavity mode with varying incidence angle inside the GaN substrate. 
The layer-resolved absorption spectra as a function of the angle of incidence for this system are reported in Fig.~\ref{fig:MQW}b,c 
and reveal that absorption occurs mostly in the doped GaN mirror. Notably, for an internal incidence angle of \dg{48} at the 
guided mode frequency, all the light is dissipated in the mirrors, leading to a minimum of zero reflectance in Fig.~\ref{fig:MQW}a.

We now turn to the system's response when the cavity is partially filled with a MQW structure (Fig.~\ref{fig:MQW}d-g). 
The superlattice is composed of twenty repetitions of a \nmetr{3} thick GaN QW, Si-doped with a concentration of 
$2 \times 10^{13}$~cm$^{-2}$ and twenty-one loss-less \nmetr{10} thick Al$_{0.26}$Ga$_{0.74}$N barriers. 
Since the guided mode is a TM mode, it naturally provides a component of the electric field along the $z$-direction 
for non-zero angles of incidence, which fulfills the ISB transition selection rule. 
In order to maximize the coupling between the ISB transition in the MQW and the cavity mode, the superlattice is placed where the $z$-component of the electric field is the largest, that is just below the Au mirror, as shown at the bottom of Fig.~\ref{fig:MQW}.
The QW dielectric tensor is modeled using a semi-classical approach \cite{Zauzny1999}. 
We selected the QW dimension and doping level in a way that it sustains a strong, collective electronic excitation known as an intersubband 
plasmon \cite{Vasanelli2016}. The components of the real part of the dielectric permittivity tensor of all the layers 
are presented in Fig.~\ref{fig:MQW}h,i. Note that while the ISB transition in the quantum well has a resonance frequency 
of $\sim \wavenumber{1600}$, the Coulomb interaction between the QW electrons results in an ENZ mode at 
$\sim \wavenumber{2030}$, marked by the dotted line in Fig.~\ref{fig:MQW}i. The ENZ mode dominates the optical 
response of the QWs. 

We show in Fig.~\ref{fig:MQW}d the calculated reflectance $\R^{p}$ of the cavity containing the MQW. The white dotted line 
shows the ENZ frequency. A clear anti-crossing can be seen between the cavity mode and the ENZ resonance, which is characteristic 
of the strong light-matter coupling regime, resulting in two polariton branches. The minimal separation between the two branches amounts to a vacuum 
Rabi splitting $2\Omega_{R} = \wavenumber{200}$. 
The layer-resolved absorption spectra as a function of the angle of incidence are shown in Fig.~\ref{fig:MQW}e-g for the GaN mirror, 
the MQW superlattice, and the Au mirror, respectively. 
When the cavity mode is far detuned from the ENZ frequency, the absorption occurs mostly in the two mirrors, and especially 
in the GaN mirror, as for the empty cavity case. The situation changes dramatically when the cavity mode is tuned near the ENZ frequency. 
The absorption then mostly occurs in the MQW. It is however important to note that the absorption is maximal at the frequencies of the two
polaritons and not at the ENZ frequency, as it would occur in the case of a weak coupling between the cavity mode 
and the QW resonance. For an internal incidence angle of \dg{48}, the maximum absorption inside the MQW stack is now 
0.5 at each of the frequencies of the two polaritons. 

The algorithm presented here thus allows to directly 
calculate the light absorption in the complex case of a doped MQW superlattice strongly coupled to a cavity mode. 
In addition to the known features, such as the avoided crossing in the angle-dependent reflectance spectrum, we can directly 
calculate the absorption inside the MQW active region, which can be usefully linked to the detected photocurrent 
in the perspective of using such structures in photodetector devices 
\cite{Liu1997,Dupont2003,Vigneron2019}.

\section{Discussion}
We have presented three nanophotonic devices based on anisotropic multilayer structures made from metals, polar dielectrics, and TMDC monolayers, covering incident wavelengths from the far-IR up to the visible. Our formalism allows to calculate the transmittance and absorption in any layer, giving unprecedented insight into the physics of light propagation in anisotropic, and even birefringent, stratified systems. 

In recent years, the field of nanophotonics has intensely investigated the optical response of 2D heterostructures. A particularly thriving subject has been polaritonic excitations, which can be supported by a broad variety of systems including slabs of metals, doped semiconductors, polar dielectrics, 2D materials such as TMDC monolayers and their stratified heterostructures \cite{Xia2014,Caldwell2015a,Low2017}. Key features of polaritons for nanophotonic technologies are their high spatial confinement and field enhancement, which are driven by the particular design of the multilayer materials, stacking order, and layer thicknesses. Potential applications of such systems range from sensing \cite{Neuner2010,Berte2018} and solar cells \cite{Luk2014}, over optoelectronic devices \cite{Freitag2013} and beam manipulation via metamaterials \cite{Zeng2013}, to waveguiding \cite{Liew2010,Passler2019a}, and ultrafast optical components \cite{Ni2016,Passler2019}. However, due to the lack of a general formalism, the optical response of these polaritonic multilayer systems is often either approximated by effective, isotropic permittivity models \cite{Liu2014,Li2014,Huber2017}, or described by specifically derived formulas \cite{Collett1971,GiaRusso1973,Schwelb1986,Ciumac1994}. Our generalized formalism allows for a precise, layer-resolved study that includes any isotropic, anisotropic or even birefringent response of any number of layers, and thus holds great potential for the prediction and analysis of polariton modes in stratified heterostructures.

This is especially relevant for systems where one or more materials feature anisotropic permittivity with spectral regions of hyperbolicity where the principle real permittivities have opposite signs, such as hBN or \MoO. Recently, these hyperbolic materials have attracted increasing interest \cite{Li2015,Ma2018,Dai2018,Ratchford2019} due to the existence of hyperbolic polaritons, featuring novel properties for nanophotonic applications such as subdiffraction imaging and hyperlensing \cite{Dai2015,Ferrari2015}. Because of the strong anisotropy of these systems, an effective permittivity approach is not purposeful. Here, our formalism provides the essential theoretical framework that is necessary to model, predict and analyze the optical response of such hyperbolic heterostructures, as we have discussed exemplarily for a \MoO/AlN/SiC system (Fig.~\ref{fig:MoO3}). Providing the full layer-resolved information about the field distribution and power flow of the excited polaritons, our method allows to readily and concisely model and study hyperbolic polariton modes in any anisotropic stratified heterostructure.

Furthermore, the layer-resolved absorption formalism provides a description for designing optoelectronic devices such as detectors, for which the photoresponse is linked to the light absorption solely in the active region of the device. In the case of the MQW system, optimizing the overall light absorption by minimizing both the reflectance and transmittance of the system would not be sufficient to model the performances of a photodetector in the strong light-matter coupling regime \cite{Dupont2003,Vigneron2019}. Unfortunately, these are the only quantities that can be probed in a reflectance experiment. Calculating the layer-resolved absorption in the structure allows to optimize the cavity and MQW design, aiming at minimizing the light absorption in the cavity mirrors while maximizing the absorption in the MQW. This is well exemplified in Fig \ref{fig:MQW} e,f where we can see that the light is preferably dissipated in either the doped GaN mirror or the MQW superlattice depending on the detuning between the cavity mode and the QW resonance. This behavior cannot be deduced only from the reflectance measurement simulated in Fig.~\ref{fig:MQW}d. Fitting experimental reflectance data using our formalism \cite{Passler2017a} would allow to retrieve the amount of light dissipated inside the active region only from the experimentally observable quantities, and to estimate figures of merit such as the quantum efficiency of the device. The present method thus provides a convenient way to model and optimize complex, optically anisotropic heterostructures for optoelectronic devices.

\section{Conclusion}
%In this work, we derive explicit expressions for the layer-resolved transmittance and absorption in stratified heterostructures of arbitrarily anisotropic, birefringent, and absorbing media, using the electric field distribution provided by our previous transfer matrix formalism \cite{Passler2017a,Passler2019b}. Our algorithm is numerically stable, yields continuous solutions, and can be easily implemented in a computer program \cite{}, enabling a robust and consistent framework that is capable of treating light of any polarization impinging at any incident angle onto any number of arbitrarily anisotropic, birefringent, and absorbing layers.
In this work, we have derived explicit expressions for the calculation of the layer-resolved transmittance and absorption of light propagating in arbitrarily anisotropic, birefringent, and absorbing multilayer media. The algorithm relies on the electric field distribution computed from a $4 \times 4$ transfer matrix formalism \cite{Passler2017a, Passler2019b}, yielding a robust and consistent framework for light-matter interaction in stratified systems of arbitrary permittivity, which is implemented in an open access computer program \cite{Passler2020,Jeannin2020}. As case studies, we applied the algorithm to simulations of three nanophotonic device structures featuring hyperbolic phonon polaritons in a polar dielectric heterostructure, \MoS~excitons in a Fabry-P\'{e}rot cavity, and ENZ resonances in a cavity-coupled multi-quantum-well, where we observed azimuth-dependent hyperbolic polariton tunneling, layer-selective exciton absorption, and strong coupling between ENZ and cavity modes. Allowing for a detailed analysis of the layer-resolved electric field distribution, transmittance, and absorption of light in any multilayer system, our algorithm holds great potential for the prediction of nanophotonic light-matter interactions in arbitrarily anisotropic stratified heterostructures.

\section{Acknowledgments} 
%\acknowledgments
We thank M. Wolf, S. Wasserroth and R. Ernstorfer (FHI Berlin) for careful reading of the manuscript and M. Wolf and the Max Planck Society for supporting this work.

\bibliographystyle{apsrev4-1}
\bibliography{layer_resolved_absorption} 

\end{document}